# Design and Construction of a Wireless Robot that Simulates Head Movements in Cone Beam Computed Tomography Imaging

R. Baghbani[1], M. Ashoorirad[1], F. Salemi[2], Med Amine Laribi[3], M. Mostafapoor[2]

[1] Biomedical Engineering Department, Hamedan University of Technology, Hamedan, Iran.

[2] Department of Oral and Maxillofacial Radiology, Dental School, Hamadan University of Medical Sciences, Hamadan, Iran.

[3] Department of GMSC, Pprime Institute CNRS, ENSMA, UPR 3346, University of Poitiers, Poitiers, France.

**SUMMARY**

One of the major challenges in the science of maxillofacial radiology imaging is the various artifacts created in images taken by cone beam computed tomography (CBCT) imaging systems. Among these artifacts, motion artifact which is created by the patient has adverse effects on image quality. This paper aims to present the design and development of a cable-driven parallel robot to study the effect of different types of head movements during CBCT imaging. The proposed robot allows a dry skull to execute motions with 3-degrees of freedom (3-DoF) during imaging in synchronous manner with the radiation beam. The kinematic model of the robot is presented to investigate the correlation between the amount of motion and the pulse width applied to dc motors, and the relevant relationships are described. This robot can be controlled by the user through a smartphone or laptop by entering the input password wirelessly via a Wi-Fi connection; thus there is not require an additional software to be installed on the smartphone or laptop. Using wireless communication prevents the user from being exposed to harmful radiation during robot driving and functioning. The results show that the designed robot has a reproducibility of over 95% in performing various movements.

1. Introduction

One of the challenges in preparing images of the head and neck is the movement of patients' heads during the preparation of images. Excessive imaging time of CBCT images (from 5-40 sec) makes patients, especially children, unable to remain still during the imaging [1]. On the other hand, patients with systemic diseases like Parkinson's should be considered during preparing images, because the patient's movements leads to the creation of movement artifacts [2]. To examine head movements in CBCT imaging, it is necessary to take an image while performing the desired head movement and compare it with the reference image (the motionless image). To emulate head movements, a robot is needed that can execute the desired movements in a controlled manner. Various studies have been conducted to investigate the effect of cranial movements on CBCT images. Naderi et.al, at the University of Florence in Italy, placed a dry skull on a long wooden bar and attached the other end to a system to create a variety of motions that transmit motion through a wooden bar to the dry skull[3,4]. This system has limited head movements. It is also necessary for the operator to be present at the imaging site in order to start and create different movements and as a result, he would be exposed to radiation. In another study, Robinson et.al, at the University of Aarhus in Denmark used a hexapod robot to create various movements[5,6]. This robot is programmable and is able to create different types of movements hydraulically and pneumatically. The main disadvantage of this system is its expensive cost. Also, the presence of metal parts in the imaging area can be problematic in creating metal artifacts. In another study at Jondishapur university, Dabbaqi et.al, examined the effect of patient movement during CBCT imaging. In this study, in order to create motion during imaging, a device was made that could move in the vertical direction (y axis) by 2, 4 and 7 mm in 2 sec. This device consists of two main parts: 1- mechanical part, which includes a stepper motor, a shaft, a bush, and connections related to converting rotational motion to linear movement; 2- Control panel, which includes circuits and control keys for different operating conditions of the device [7,8]. The problem with this system is that it can only move in one direction and therefore, it is not possible to perform movements such as rotation, vibration and combination of motions together.

Cable-driven parallel robots (CDPRs) are increasingly applied in different tasks, like rescue systems, rehabilitation, and even 3-dimensional print. A typical CDPR system is formed from three parts, including a fixed platform, a mobile platform, and a number of other cables, which are wont to join the fixed platform and mobile platform. The cable length can be changed through winches actuated by motors installed in the fixed platform[9–11].

In this study, a cable-driven parallel robot is design to assist researchers in the study of motion artifacts in CBCT imaging by creating different motions in the dry skull. The prototype of robot has been developed and validated through the experimental tests.

This paper is organized as follows: Section 2 deals with the description of the mechanical and electrical components as well as kinematic model of the robot along with explanation of correlation of DC motor speed with PWM duty cycle. In section 3, description of the robot prototype along with the real tests on CBCT imaging is presented. In Section 4, The capabilities and limitations of the designed robot are discussed. The last section concludes the paper.

## 2. Materials and methods

In designing the robot, it is necessary to take into account the limitations of the construction and utilizations steps. To design a robot that is supposed to emulate and execute possible movements of the patient's head during the CBCT imaging, the following items should be considered:

1- The robot dimensions should be designed in such a way that it does not collide with the arm of the radiation tube and the detector of the CBCT imaging system during operation.

2- The constructed robot must have the ability to execute and simulate any required head movement.

3- To prevent the operator from radiation, the robot must be able to set up and control remotely.

4- Different parts of the robot should be replaced in a short time so that if a part of the robot is damaged, that part can be easily replaced.

5- The designed robot must have sufficient mechanical strength against various movements and be able to withstand the force applied to it during the use.

The robot designed in this research consists of mechanical and electrical components, each of which is discussed below.

### 2.1. Mechanical components of the robot

The various mechanical parts of the robot that are designed in the COMSOL software environment are shown in Fig. 1.

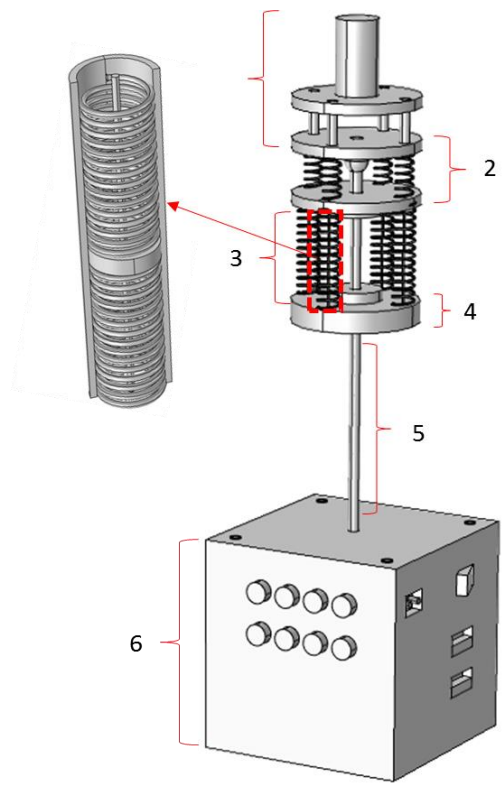

Fig. 1: Mechanical components of the head movement simulator robot of 3-dofs along with kinematic architecture and head fixation.

According to Fig. 1, the mechanical components of the robot are divided into six parts, which are described in the following. Part 1 of Fig. 1 shows the chassis and location of the dry skull. Since this part is close to the skull and in its connection, it is exposed to radiation, and therefore to prevent the metal artifact in the images [4], this part is made entirely of polyamide and Plexiglas. The dry skull mounts on a polyamide cylinder through its circular base cavity.

Part 2 of Fig. 1, with the help of a spherical joint embedded in the middle, provides displacement of the dry skull mounted on the chassis. This part also includes four suitable springs that are installed in the robot to maintain balance and return the skull to its original position. The tension of the skull is provided by four flexible cables that are connected to the lower surface of the skull chassis through the four springs.

Part 3 of Fig. 1 consists of four cylinders through which the flexible cables pass as shown in Fig. 2. The flexible cables are connected at one end to the pulley mounted on the shaft of DC motor and at the other end to the underside of the skull chassis.

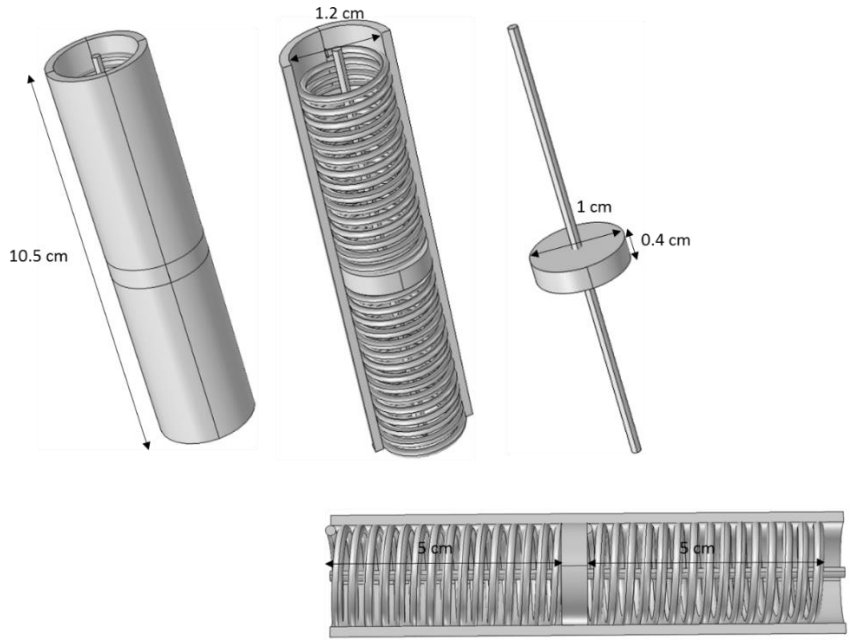

Fig. 2: Design of tension system and springs in COMSOL environment. The tension system in two different directions includes a cylinder, two springs separated by a polyamide disc and a flexible cable.

According to Fig. 2, the two springs in the middle of the cylinder are separated by a solid disk. The cylinder is also fixed from the top and bottom by the top and bottom plates. The flexible cable is firmly attached to the solid disc in the middle, making it possible for the flexible cable to be stretched up and down. The flexible cable is transferred to the upper parts after passing through the hole created in the upper spring axis of the cylinder, and from the lower part of the cylinder, after passing through the appropriate hole, it is connected to the pulley connected to the gearbox shaft until the flexible cable is pulled or released around the pulley during the motor functioning.

Part 4 of Fig. 1 contains the chassis and location of the four DC motors. According to Fig. 3, DC motors must be carefully positioned so that the rotational motion is converted to linear displacement with complete accuracy.

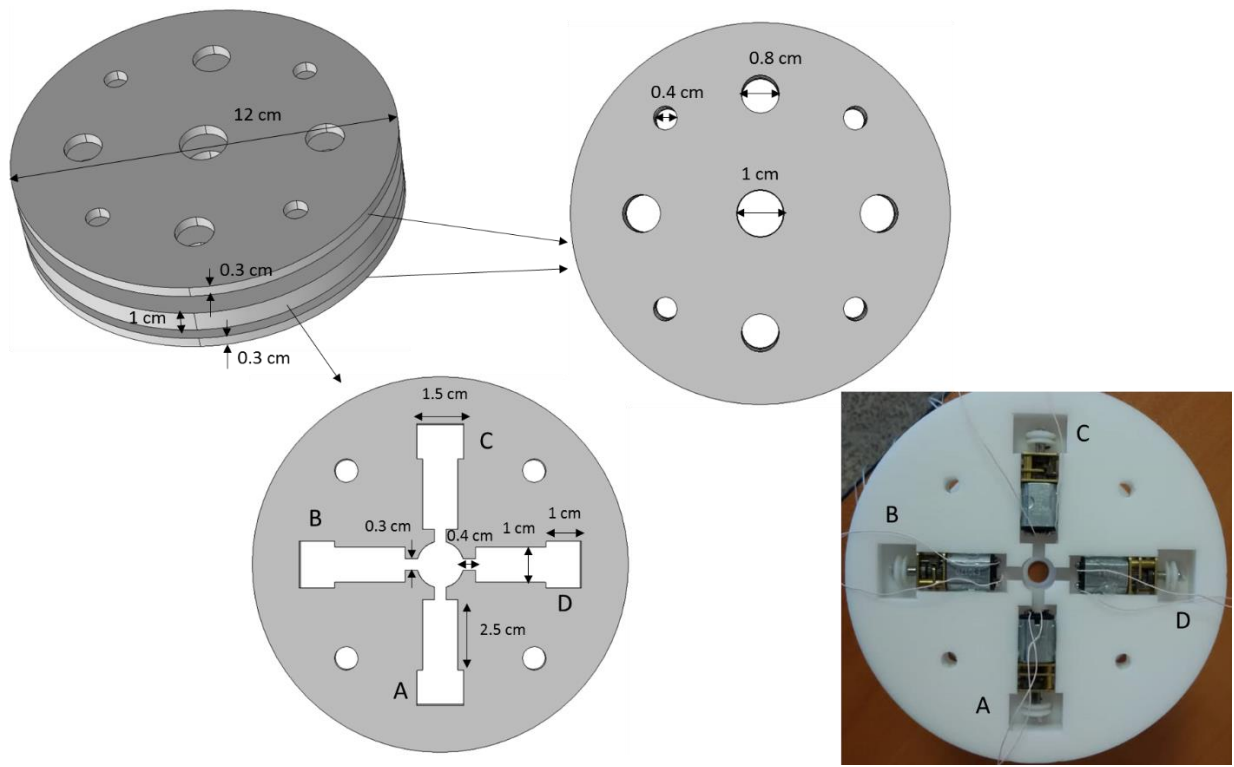

Fig. 3: (A) Chassis design and location of four DC motors in the COMSOL environment; (b) Chassis construction and placement of four DC motors in it.

As shown in Fig. 3, the power wires connected to the motors pass through the built-in slots and connect to the L298 drivers. A circular hole in the middle of the DC motor chassis is installed to pass the steel rod to transmit rotational motion.

Part 5 of Fig. 1 shows the steel rod used to transmit rotational motion to the dry skull without delay. This rod is connected from one end to the stepper motor through a suitable bushing and from the other side to the roof of part 3.

Part 6 of Fig. 1 contains the box of electronic circuits and boards. Also, in the roof part of this box, a stepper motor is firmly fixed. Manual adjustment switches, USB ports for programming of the two boards (Arduino and ESP32), and city power socket are installed on the body (part 6).

**2.2. Electrical components of the robot**

The electrical parts of the robot are in the form of a block diagram shown in Fig. 4.

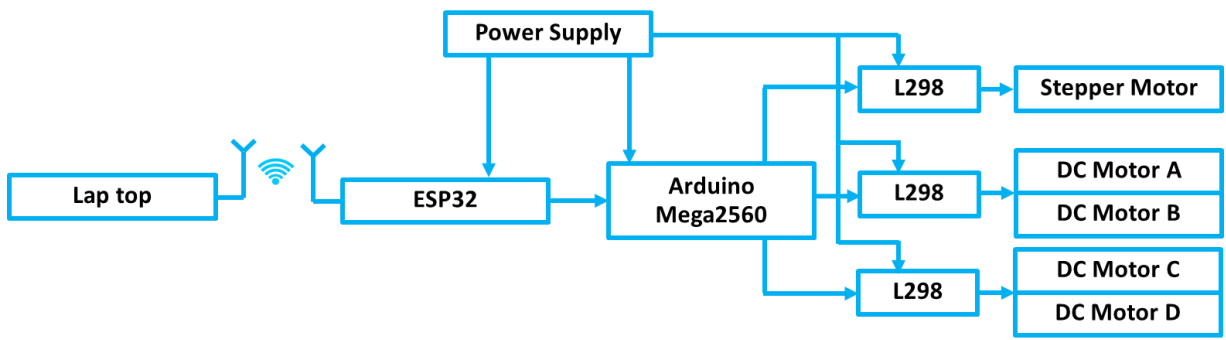

Fig.4: Block diagram of the robot's electrical components

As shown in Fig. 4, the robot is designed with four DC motors to generate linear motion. The DC motors are powered by the L298 drivers that accept digital input. The pulse width modulation (PWM) method has been used to start and control the speed of these four motors [12]. The amount of displacement is controlled by the duration of the PWM wave applied to the DC motors. For more displacements, the wave duration needs to be longer. The DC motors at idle have a rate of 3000 revolutions per minute (rpm), but the number of revolutions decreases with the application of the load. To keep the number of revolutions constant, or in other words to reduce the effect of loading, a gearbox connected to the motor shaft is used, which converts the number of revolutions to 30 rpm or half rotation per second when the applied DC voltage is 6 V. Therefore, by increasing and decreasing the PWM pulse width, the rotation speed will increase and decrease. The presence of the gearbox, in addition to drastically reducing the loading effect during start-up with the load, also acts as a brake and lock at rest, causing the dry skull to remain stationary in a certain position.

The four DC motors are used to create four basic types of motion: posterior motion, anterior motion, lateral left motion, and lateral right motion by simultaneously turning on two motors in all four motions. By starting the motors at certain times, certain combination of the movements can also be executed. The stepper motor used in this robot is used to create rotational movements (angular displacement) at different and defined angles. This motor has also been used to create vibrational motion. The step of this motor is 1.8 degrees, which is used in this design as a half-step with an angle of 0.9 degrees. Therefore, the minimum possible displacement angle in this robot is 0.9 degree. The stepper motor is connected to the Arduino via the L298 driver. Based on the user's choice of motion, the Arduino executes the algorithm required to start the stepper motor.

The main part of the robot is its CPU, which uses the Arduino Mega2560. This board is required to execute the motion algorithms written to perform the movements at a given time. Therefore, to design a movement, it is enough to write the necessary program to start DC motors at specific times and program it to the processor through the embedded USB port. As a result, different movements for dry skulls can be designed according to the need in different dental researches.

One of the important parts of the designed robot is the Wi-Fi wireless communication unit. The ESP32 board has been used for this purpose [13]. With the programming, this board is able to communicate with a smartphone or laptop via Wi-Fi as shown in Fig. 5. The importance of setting up and controlling the robot wirelessly is that the presence of the user is not required to launch the robot during imaging, and this prevents the person from being exposed to radiation.

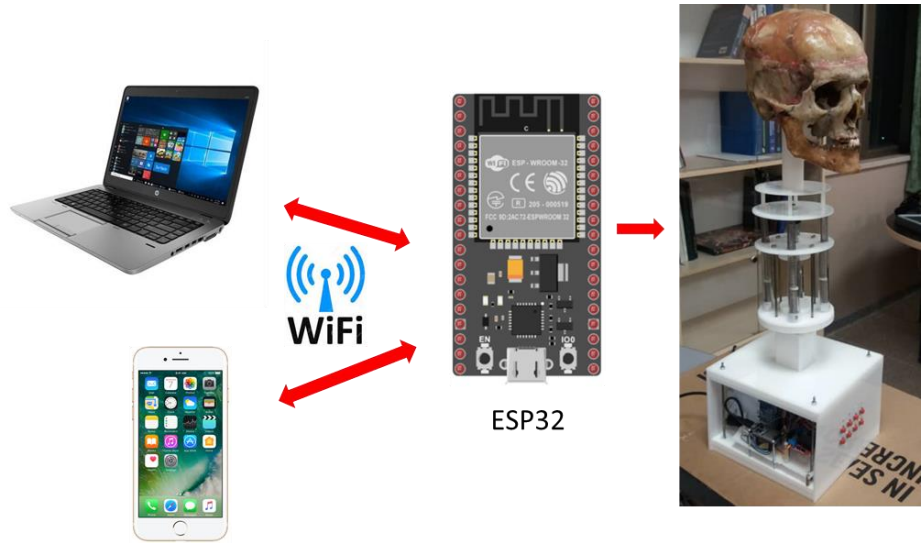

Fig. 5: The robot Wi-Fi wireless connection with a smartphone or laptop.

The feature of this type of wireless communication is that there is no need for any other application on the smartphone or laptop. As soon as the ESP32 board is launched, the corresponding Wi-Fi is included in the list of Wi-Fi of the phone or laptop, and the user communicates with the robot by selecting it and entering the intended password; then the robot control panel, which contains all the control keys for executing the different motions appears on the screen. By selecting each key by the user, the corresponding motion is performed by sending the corresponding motion code to the Arduino board. The program written for the ESP32 to design the control panel can be changed, so the user can design the number of keys and their titles as needed to appear on the screen of the phone or laptop.

To supply electricity, a switching power supply is used, which has two outputs, 5 V and 12 V. It has been used for different parts by connecting a series of several diodes in order to reduce the voltage.

**2.3. Kinematic model of the robot**

The kinematic model describes the motion of a robot mathematically without considering the forces affect the motion and engages itself with the geometric relationship between the elements. A simple kinematic model of the designed robot is shown in Fig. 6.

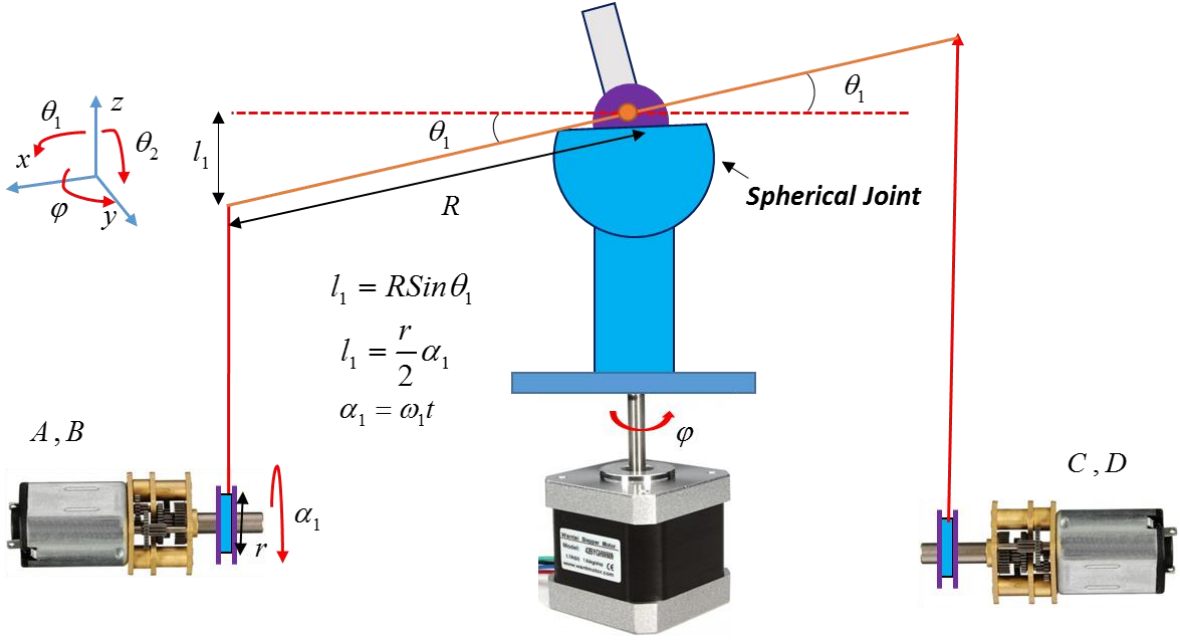

Fig. 6: A simple kinematic model for the designed cable driven robot to emulate head movement during the CBCT imaging.

As shown in Fig. 6, the angular displacement of the rotating plate connected to the spherical joint, equal to $\theta_1$(Rot(y)), results in a linear displacement in the vertical direction ($l_1$) in the x-z plane (anterior-posterior motion). This is done by pulling the flexible cable with the help of the two DC motors A, B and simultaneously releasing the flexible cable by the two DC motors C, D. For angular displacement of $\theta_2$ (Rot(x)), which consequently results in the linear displacement of $l_2$ on the y-z plane, it is necessary that the two motors A, D pull the flexible cable and the motors B, C release the flexible cable (lateral movement). The relationships governing this mechanism are given in Eqs. (1) and (2).

$$l_1 = R\sin\theta_1 = \frac{r}{2}\alpha_1$$
$$l_2 = R\sin\theta_2 = \frac{r}{2}\alpha_2 \tag{1}$$

$$\alpha_1 = \omega_1 t$$
$$\alpha_2 = \omega_2 t \tag{2}$$

In the (1), (2) relations, $\alpha_i$ is the amount of rotation of the pulley connected to the end of the motor gearbox. The diameter of the pulley, $r$, in all four dc motors is $1\ cm$. The angular velocity of the motor at the gearbox output is equal to $\omega_i$ in terms of rpm. Also, $R = 5\ cm$ is the radius of the rotating plate connected to the spherical joint. According to the (1), (2) equations, having the angular velocity, $\omega_i$, and the running time of the DC motors, $t$, the amount of angular and linear displacement, $\theta_i$ and $l_i$, could be determined. Also, a stepper motor is used to rotate the skull by φ self-rotation. The speed of the rotation depends on the amount of delay between the excitation phases and the amount of rotation depends on the number of stimuli and the step of the stepper motor, which are both completely controllable.

**2.4. Correlation of the DC motor speed with PWM duty cycle**

In order to determine the exact amount of linear displacement of the plate connected to the spherical joint, the PWM pulse width method with a frequency of 1 kHz has been used. To do this, the speed of the dc motor is measured by increasing the pulse width. The PWM pulse is applied to the input of the L298 driver and the power input of this driver is connected to a voltage of 12 V. So, for example, if the PWM pulse width is 50%, the average output voltage would be 6 V. The changes in the speed of the dc motor at the gearbox output in terms of the changes in pulse width from zero to 50% are obtained as a curve in Fig.7.

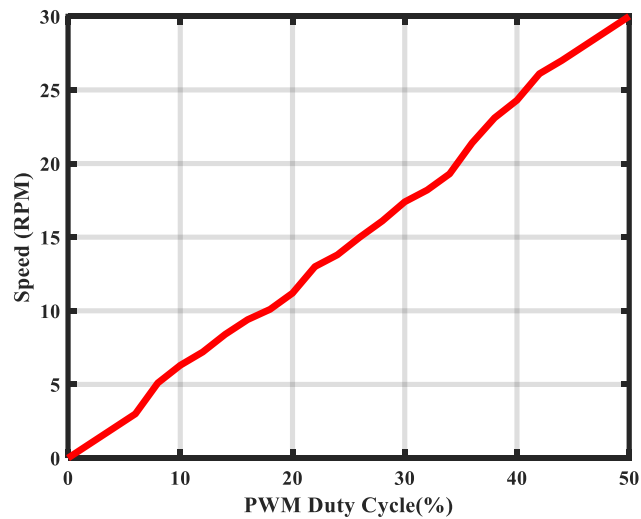

Fig. 7: Changes in the angular velocity of the DC motor at the gearbox output as the PWM pulse width increases.

Fig. 7 shows the increase in angular velocity of the DC motor with a gearbox with a conversion ratio of 100:1 in terms of pulse width. Although the motor output is measured at no load, the load effect during starting with load will be negligible due to the use of the gearbox with large conversion ratio. The relationship between the angular velocity and the width of the PWM pulse using linear regression is approximated as follows:

$$\omega(rpm) = 0.609\, PWM\,(\%) \qquad (3)$$

By substituting Eq. (3) for Eqs.(1) and (2), we can write:

$$\theta_1 = Sin^{-1}(\frac{rt}{2R}0.609 PWM_1)$$

$$\theta_2 = Sin^{-1}(\frac{rt}{2R}0.609 PWM_2) \qquad (4)$$

The choice of PWM duty cycle and duration $t$ is easily possible with the Arduino board processor. As a result, by determining $\omega_i$, the values of $\alpha_i$ and $l_i$ can be controlled.

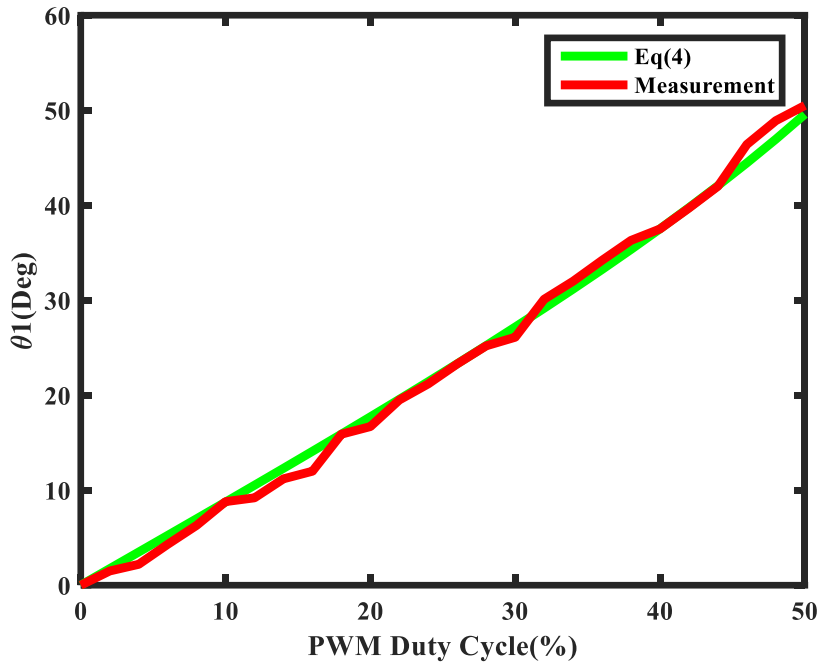

Fig .8: Comparison of the $\theta_1$ values obtained from Eq. (4) and the measured values related to changing the PWM at the t=250 ms, r=1 cm, and R=5 cm.

For example, the variation curves of $\theta_1$ using Eq. (4) and measured values of $\theta_1$ in terms of the PWM changes are plotted in Fig. 8. As shown in Fig. 8, there is a good match between the output of the kinematic model and the actual output of the robot. The normal error between the values of these two curves in Fig. 8 is calculated using Eq. (5) as the follow.

$$Normerror = 100(\frac{\sqrt{\sum_{i=1}^{N}(\theta_i - \bar{\theta})^2}}{\sqrt{\sum_{i=1}^{N}(\bar{\theta})^2}}) \qquad (5)$$

where, $\bar{\theta}$ is the measured average of $\theta_1$ and $\theta_i$ is obtained value of $\theta_1$ using Eq. (4), and N=5 is a number of measurements. Error value of e = 4% is obtained.

### 3. Results

After designing and manufacturing various mechanical and electrical parts of the robot, the robot components are assembled and finally the robot is made. The different parts of the robot have been made in such a way that their connections are performed easily and the possibility of replacing the components can be done quickly. The initial program was programmed for four different motions at different speeds on the Arduino board processor. The ESP32 board was programmed to for Wi-Fi connection. As soon as the robot is turned on, according to the program written for the ESP32, the Wi-Fi icon related to the robot is appeared on smartphone or laptop near the robot. Fig. 8 shows the control panel page that has been designed for the robot used in this research.

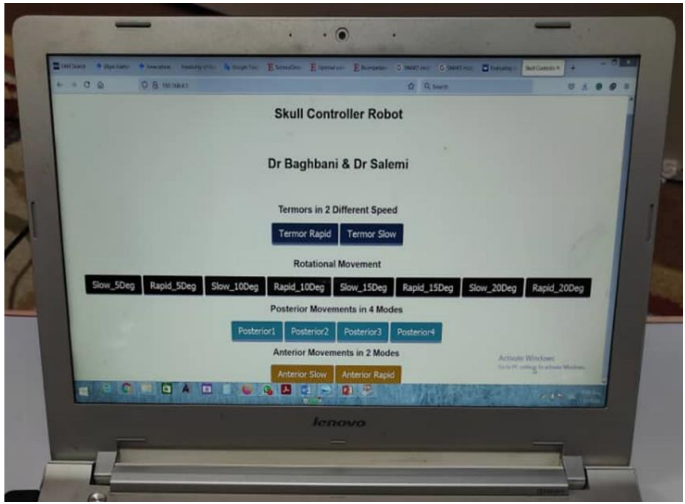 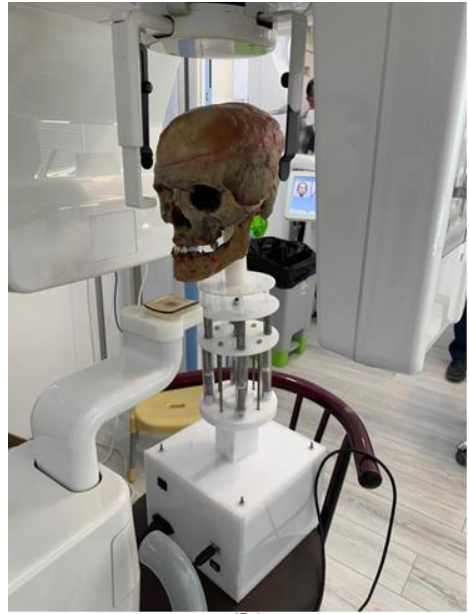

Fig. 8: Setting up the robot in the CBCT imaging system. (a) The control panel page after connecting the laptop to the robot via Wi-Fi (b) Placing the robot in the CBCT imaging device.

According to Fig. 8 (a), four default motions have been considered for the robot, which are:

1- Tremor motion of 5 degrees at two different speeds.

2- Rotational motion at four different degrees of 5, 10, 15 and 20 with two different adjustable speeds.

3. Posterior movement with four different displacements of 5, 10, 15 and 20 mm.

4- Anterior movement with two different displacements of 5 and 10 mm at two different speeds.

The position of the robot to perform the motions during CBCT imaging is shown in Fig. 8 (b). As mentioned, these movements are defined by default for the robot, and it is possible to combine different movements as needed by reprogramming the processor. For example, performing simultaneous anterior motion with self-rotation movement can create a combination of rotational and anterior motion. An important point in the robot motions planning is that in order to correctly perform the linear motions (anterior, posterior, right lateral, and left lateral movements) and their combinations, one phase of the stepper motor is stimulated; then, the rotational movement is completely locked. This helps a lot to perform the mentioned linear movements accurately. As soon as the aforementioned linear movements are completed, the single-phase excitation of the stepper motor is stopped so that the phase winding is not damaged by heat. To evaluate and analyze images with different motion artifacts, in each type of motion, a reference image is first taken in which the robot does not make any movement. Acrylic balls mounted on dry skull at anterior and posterior region of maxillary and mandibular alveolar ridge have also been used to investigate the effect of various motion artifacts on imaging quality. The distance between acrylic spheres on both the skull and CBCT images was measured and compared. The distance between two acrylic spheres in images with motion artifacts relative to the motionless reference image is used as an indicator to quantify the amount and intensity of the artifact[14].

Imaging of the dry skull in the non-moving position and in moving the head was performed by two different CBCT imaging systems. These devices have been used in previous studies[15–17]. Table 1 shows the specifications of these two CBCT systems.

Table 1: The specifications of the two CBCT imaging systems used in this research.

| CBCT imaging systems | T (sec) | mA | kVP | Field of view (FOV) | Voxel size |
|---|---|---|---|---|---|
| Cranex 3D device (Soredex, Helsinki, Finland) | 14.2 | 10 | 90 | 6 cm × 8 cm | 200 µm |
| Kodak device (Care Steam, France) | 12 | 10 | 90 | 8 cm × 8 cm | 200 µm |

To evaluate the robot's performance, CBCT imaging was performed in motionless state, rotational motion of 5 degrees, rotational motion of 10 degrees, linear posterior motion of 5 mm. The imaging results in these cases using the Cranex 3D device are shown in Fig. 9.

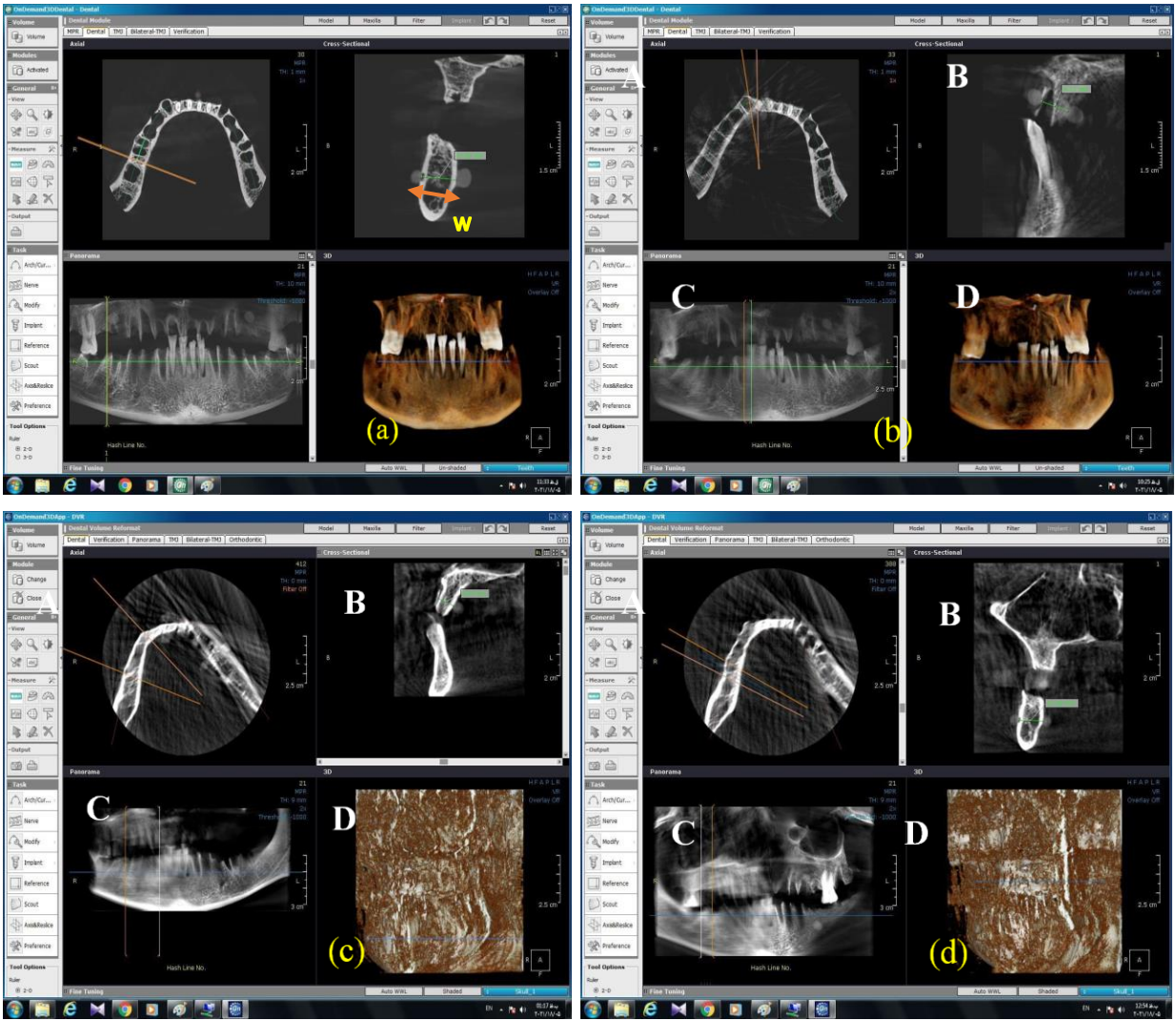

Fig. 9: Cranex 3D imaging and measurement of the posterior mandibular area from CBCT axial images (A): Axial section (B): Cross section (C): Reconstructed panoramic image (D): Reconstructed 3D image. (a) Reference image (without motion) (b) Image with a rotational motion of 5 degrees (c) Image with a rotational motion of 10 degrees (d) Posterior motion of 7 mm.

Before and after making various head movements, CBCT images were prepared and the resulting images were stored in Ondemand software (Cybermed, Ceul, Korea). According to Fig. 9, and the figure corresponding to the axial section (A), the image of the 10-degree rotational motion in part (c) has created more artifacts than the 5-degree rotational motion in part (b). Also, in the image related to the posterior movement of 7 mm (part (d)), the created artifact is much more than the artifact caused by the 10-degree rotational movement.

To study the reproducibility of performing various movements by the robot, the different motions were repeated 5 times and the normal error for them was calculated according to Table 2.

Table 2: The normal errors between the motionless (reference) CBCT image and the CBCT images with different motion artifacts along with the distance between the two acrylic balls on the dry skull in different CBCT images (W).

|  | motionless | self-rotation motion ($\varphi=5°$) | self-rotation motion ($\varphi=10°$) | posterior motion of 5 mm | posterior motion of 10 mm | Anterior motion of 5 mm | Anterior motion of 10 mm |
|---|---|---|---|---|---|---|---|
| Normal error(%) | 0 | 0 | 0 | 3.4 | 4.5 | 3.6 | 4.7 |
| W (mm) | 9.72 | 9.35 | 9.18 | 9.32 | 9.44 | 8.92 | 8.56 |

According to Table 2, the normal error obtained for these six different movements is less than 5%. For the self-rotation motion, the obtained error is, which indicates that the reproducibility of this motion is 100%.

The quality of the images can be examined in terms of the distance between the two acrylic balls placed on the dry skull at a certain distance from each other. For example, according to Table 2, the distance between the two balls for six different movements (W in Fig. 9(a)), along with the value of the motionless image is given. These distances are measured by the CBCT device software itself. The values obtained for W show that the distance factor between two acrylic balls can be used to quantify the motion effect.

### 4. Discussion

CBCT imaging is now considered by patients and dentists and has found many applications in implant therapy, endodontics, oral and maxillofacial surgeries, airway evaluation, orthodontic treatment and evaluation of pathological lesions.

One of the challenges in preparing images of the head and neck is the movement of patients' heads during the preparation of images. Excessive CBCT imaging time prevents patients, especially children, from remaining silent during imaging. Therefore, the movement of the patient's head and neck when preparing CBCT images leads to the creation of movement artifacts.

The patient's movement artifacts are seen as straight lines or streaks in the images. If the patient's movements are voluntary, cooperation can be obtained by justifying the patient and reminding him/her of the importance of staying still during imaging. In the case of patients who are unable to cooperate or have involuntary movements, short scan times should be used or the status of patients who are unable to cooperate should be stabilized using fixation devices. Of course, if there is a lot of movement, these errors cannot be corrected using computer software; This is because the image is severely distorted as the voxels move during imaging, and the imaging must be repeated. However, with the advancement of technology, it is possible to improve a variety of artifacts in CBCT imaging. Motion artifacts are more important because they occur on the patient side.

In this research, in order to investigate the effect of different types of head movements during imaging that lead to motion artifacts, a robot has been designed and implemented that is able to execute different motions on a dry skull using four DC motors and a stepper motor. Therefore, the movements created during CBCT imaging are completely controlled, which will investigate in more details in our future study. Therefore, with this robot, it is possible to investigate the main factors of motion artifacts in images. One of the main features of this robot is that it is universal, which means that with programming, it is possible to create a combination of different movements. To check for any requested motion, the robot is programmed via the built-in USB port. In order to prevent the user from being exposed to radiation, the robot is set up wirelessly using the Wi-Fi protocol. The basis of wireless communication is the ESP32 board, which its programming can also be changed. One of the important features of this board is the ability to change its program; as a result, the control screen that appears on the monitor of a laptop or smartphone is changed based on the types of requested motions. By launching this board and connecting to a smartphone or laptop, the control panel, which includes a variety of keys to execute various motions, appears on the monitor screen of the phone or laptop. So there will be no need for another application to be installed on the smartphone or laptop. Simultaneously, several intelligent systems can control the robot and launch the desired motions. The rotational motion resolution of this robot is less than one degree and the linear motion is less than two millimeters. All the parts used in this robot are replaceable and at any stage when there is a problem for different parts, it can be replaced in a short time.

One of the effective methods for evaluating the quality of images with motion artifacts is to measure the distance between two specific points in the image with artifacts relative to the distance of the same two points in the reference image[14]. These points can be determined using materials that have no effect on the image. For example, acrylic materials would be a good choice for this purpose. Since the reproducibility of performing various motions by the robot is close to 100% (Normal error<5%), so the effect of different types of motions can be studied by measuring the distance between two points created by acrylic materials.

The robot designed in this research has advantages over the studies done in the past. Ability to program multiple to perform different movements according to need, set up and remote control of the robot to create different movements and similar to real movements, and a combination of basic movements are the important features of the robot constructed in this research. Naderi et.al, at the University of Florence in Italy, designed and built a robot that could only create rotational and linear motions in one direction. In a study conducted by Dabbaqi et.al, the dry skull was able to move in one direction to create a unidirectional linear and rotational motions. In addition, it was not possible to combine motions to create complex head movements in this robot; Also, the robot is set up and controlled manually and with a wire connection. The hexapod robot was used by Robbins et.al, at the University of Aarhus in Denmark to create various dry skull motions. Although the robot was designed for a variety of other purposes, it was capable of moving the dry skull in different directions. The robot moved its arms hydraulically and moved the object on its body. The main disadvantage is the use of metal arms, which by being in the imaging area may cause artifacts in the image[18–20]. Also, the high cost of this robot and its unavailability for dental researchers and radiologists are other its disadvantages.

## 5. Conclusion

One of the main challenges of CBCT imaging is the patient's movement during imaging and the creation of motion artifacts that greatly reduce the quality of the images and in many cases it is necessary to repeat the imaging. In addition to incurring additional costs, this causes the patient to be exposed to radiation for a longer period of time. In this research, to study the effect of different types of head movements and their amount and speed in the artifact created in CBCT images, a robot has been designed and built that is able to remotely perform various movements of the dry skull during imaging. One of the features of this robot is its wireless connection via Wi-Fi to smart systems such as laptops and smartphones. By connecting to Wi-Fi, the robot control screen appears on the screen of the smartphone or laptop and without the need for additional program to launch the robot, the user can control it and the desired movements are performed by the robot during imaging.

**Acknowledgments:** This work was financially supported by Department of Oral and Maxillofacial Radiology, Dental School, Hamadan University of Medical Sciences, Hamadan, Iran, through the financial agreement number 9911288440.